\documentclass[prl,twocolumn,amsmath,amssymb,amsbsy,superscriptaddress,floatfix,nobalancelastpage]{revtex4-1}

\usepackage{graphicx,bm,times}
\usepackage{color}

\begin{document}

\title{Cascade of field-induced magnetic transitions in a frustrated antiferromagnetic metal-\\
possible experimental signature of a magnetic supersolid phase}


\author{A. I. Coldea}\email[corresponding author:]{amalia.coldea@physics.ox.ac.uk}
\affiliation{Clarendon Laboratory, Department of Physics,
University of Oxford, Parks Road, Oxford OX1 3PU, U.K.}
\affiliation{H. H. Wills Physics Laboratory, University of Bristol,
Tyndall Avenue, BS8 1TL, United Kingdom}

\author{L. Seabra}
\affiliation{Department of Physics, Technion --- Israel Institute of Technology, Haifa 32000, Israel}
\affiliation{Max-Planck-Institut f\"ur Physik komplexer Systeme, 01187 Dresden, Germany}
\affiliation{H. H. Wills Physics Laboratory, University of Bristol,
Tyndall Avenue, BS8 1TL, United Kingdom}

\author{A. McCollam}
\affiliation{High Field Magnet Laboratory, IMM, Radboud University
Nijmegen, 6525 ED Nijmegen, The Netherlands}

\author{A. Carrington}
\affiliation{H. H. Wills Physics Laboratory, University of Bristol,
Tyndall Avenue, BS8 1TL, United Kingdom}

\author{L. Malone}
\affiliation{H. H. Wills Physics Laboratory, University of Bristol,
Tyndall Avenue, BS8 1TL, United Kingdom}

\author{A. F. Bangura}
\affiliation{Max-Planck-Institut fur Festkorperforschung,
Heisenbergstr. 1, 70569 Stuttgart, Germany}
\affiliation{RIKEN(The Institute of Physical and Chemical Research),
Wako, Saitama 351-0198, Japan}
\affiliation{H. H. Wills Physics Laboratory, University of Bristol,
Tyndall Avenue, BS8 1TL, United Kingdom}

\author{D. Vignolles}
\affiliation{Laboratoire National des Champs Magn\'{e}tiques Intenses (CNRS), Toulouse, France}

\author{P. G.{van Rhee}}
\affiliation{High Field Magnet Laboratory, IMM, Radboud University
Nijmegen, 6525 ED Nijmegen, The Netherlands}

\author{R. D. McDonald}
\affiliation{National High Magnetic Field Laboratory, Los Alamos National Laboratory,
MS E536, Los Alamos, New Mexico 87545, USA}

\author{T. S\"{o}rgel}
\affiliation{Max-Planck-Institut fur Festkorperforschung,
Heisenbergstr. 1, 70569 Stuttgart, Germany}

\author{M. Jansen}
\affiliation{Max-Planck-Institut fur Festkorperforschung,
Heisenbergstr. 1, 70569 Stuttgart, Germany}

\author{N. Shannon}
\affiliation{Okinawa Institute of Science and Technology,
1919-1 Tancha, Onna-son, Kunigami, Okinawa 904-0495, Japan}
\affiliation{Clarendon Laboratory, Department of Physics,
University of Oxford, Parks Road, Oxford OX1 3PU, U.K.}
\affiliation{H. H. Wills Physics Laboratory, University of Bristol,
Tyndall Avenue, BS8 1TL, United Kingdom}

\author{R. Coldea}
\affiliation{Clarendon Laboratory, Department of Physics,
University of Oxford, Parks Road, Oxford OX1 3PU, U.K.}
\affiliation{H. H. Wills Physics Laboratory, University of Bristol,
Tyndall Avenue, BS8 1TL, United Kingdom}


\begin{abstract}

Frustrated magnets
can exhibit many novel forms of order when exposed to high
magnetic fields, however,
much less is known about materials where frustration occurs in the
presence of itinerant electrons.
Here we report thermodynamic and transport measurements on
micron-sized single crystals of the triangular-lattice metallic
antiferromagnet $2H$-AgNiO$_2$, in magnetic fields of up to 90~T
and temperatures down to 0.35~K.
We observe a cascade of magnetic phase transitions at 13.5, 20, 28
and 39~T in fields applied along the easy axis, and we combine
magnetic torque, specific heat and transport data to construct the
field-temperature phase diagram.
The results are discussed in the context of a frustrated easy-axis
Heisenberg model for the localized moments where intermediate
applied magnetic fields are predicted to stabilize a magnetic
supersolid phase. Deviations in the measured phase diagram from
this model predictions are attributed to the role played by the
itinerant electrons.
\end{abstract}



\maketitle

Frustrated magnets have proved a rich source of novel magnetic
ground states such as spin liquids on triangular lattices
and spin ices on pyrochlore lattices \cite{Balents2010}.
Another intriguing possibility, rooted in work on superfluid
Helium, is that a frustrated magnet might realise a magnetic
analogue of a supersolid, in which broken translational symmetry
co-exists with a superfluid order parameter
\cite{Matsuda1970,Liu1973}.
Most experimental and theoretical studies of frustrated magnetism
have focused on insulating systems, however, there are interesting
examples of phenomena in metallic systems where frustration is
believed to play a crucial role, such as heavy fermion physics in
the spinel LiV$_2$O$_4$ \cite{LiV2O4}, and the metallic
spin-liquid state in the pyrochlore Pr$_2$Ir$_2$O$_7$
\cite{Nakatsuji2006}.
In this context,
the layered delafossites AgNiO$_2$ \cite{growth} and Ag$_2$NiO$_2$ \cite{Ag2NiO2}
provide model systems for studying the interplay of
metallic electrons and local-moment magnetism on a geometrically-frustrated lattice.


%
Detailed structural studies on the hexagonal, $2H$-polytype, AgNiO$_2$
reveal a charge ordering transition at 365~K, below which one-third of the Ni ions form
a triangular lattice of localized Ni$^{2+}$ ($S=1$) magnetic moments,
while the remaining Ni$^{3.5+}$
ions form a honeycomb network of itinerant paramagnetic sites 
\cite{Wawrzynska2007,Wawrzynska2008,Mazin2007,Pascut2011}.
The localized Ni$^{2+}$ spins order magnetically below
$T_N=19.5$~K into a collinear antiferromagnetic structure of
alternating ferromagnetic stripes in the triangular plane, with
spins aligned along the $c$-axis, while the itinerant Ni$^{3.5+}$
sites remain paramagnetic \cite{Wawrzynska2007,Wawrzynska2008}.
Bandstructure calculations suggest that the Ag {\it sp} band is
entirely above the Fermi level and that the structure can be
visualized as a magnetic insulator formed by Ni$^{2+}$ (like NiO)
with a strong tendency to magnetic order, superimposed on a
Ni$^{3.5+}$ metal \cite{Wawrzynska2007}.

Here we report thermodynamic and transport measurements on
micron-size single crystals of 2$H$-AgNiO$_2$ in high magnetic
fields of up to 90~T applied along the $c$-axis. We observe a
complex cascade of magnetic phase transitions, and combine
magnetic torque, heat capacity and transport measurements to
construct the field-temperature phase diagram.
Experimental data are compared with the predictions of a
frustrated easy-axis Heisenberg model for the localized Ni$^{2+}$
moments, which predicts a field-driven phase transition from the
collinear antiferromagnet (CAF) into a magnetic supersolid (SS),
which can be viewed as a Bose condensate of magnons of the CAF
phase. Deviations from this model compared to the measured phase
diagram are attributed to the role played by the itinerant
electrons.

\begin{figure}[h!]
\centering
\includegraphics[width=8.5cm,clip=true]{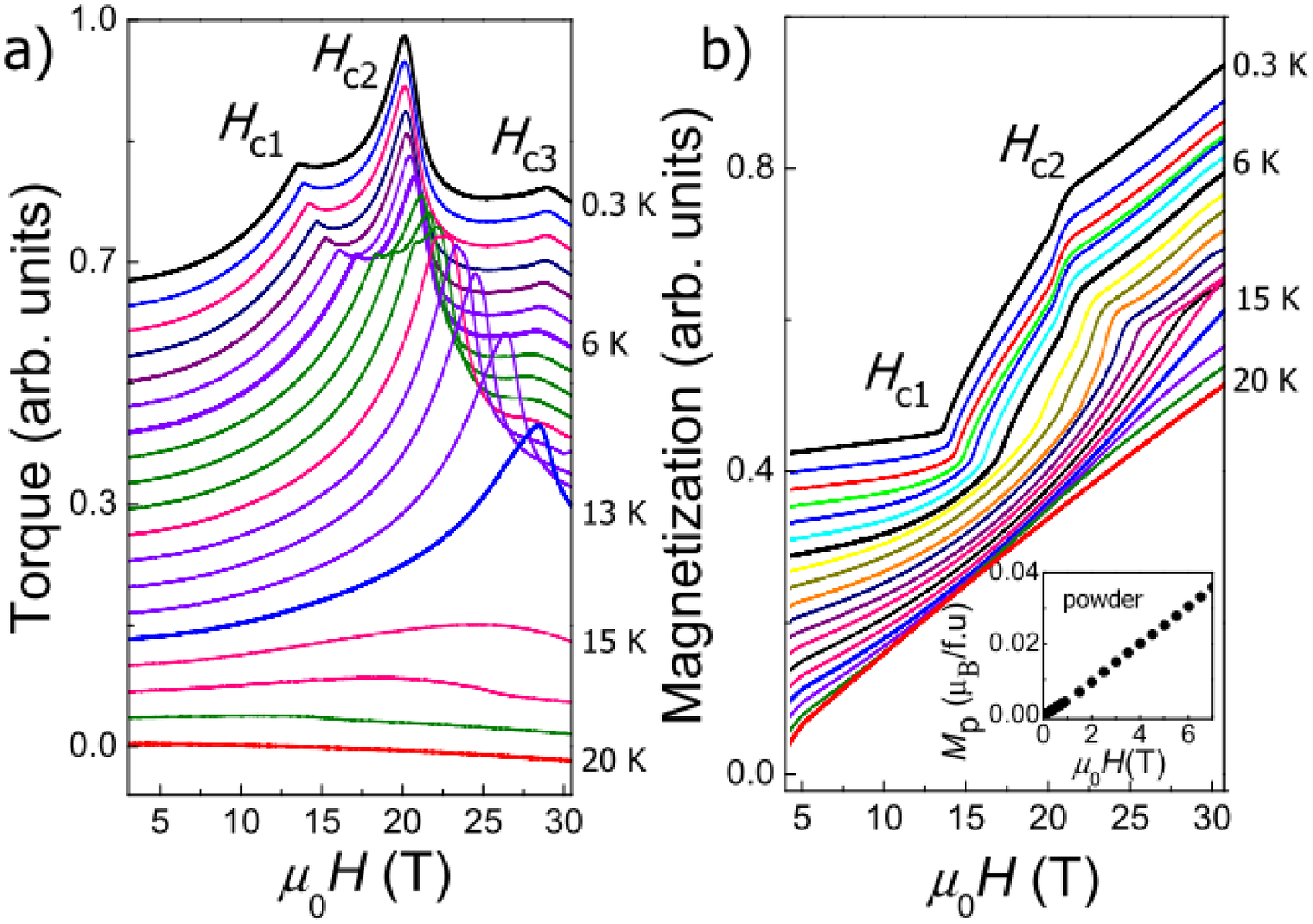}
\includegraphics[width=8.5cm,clip=true]{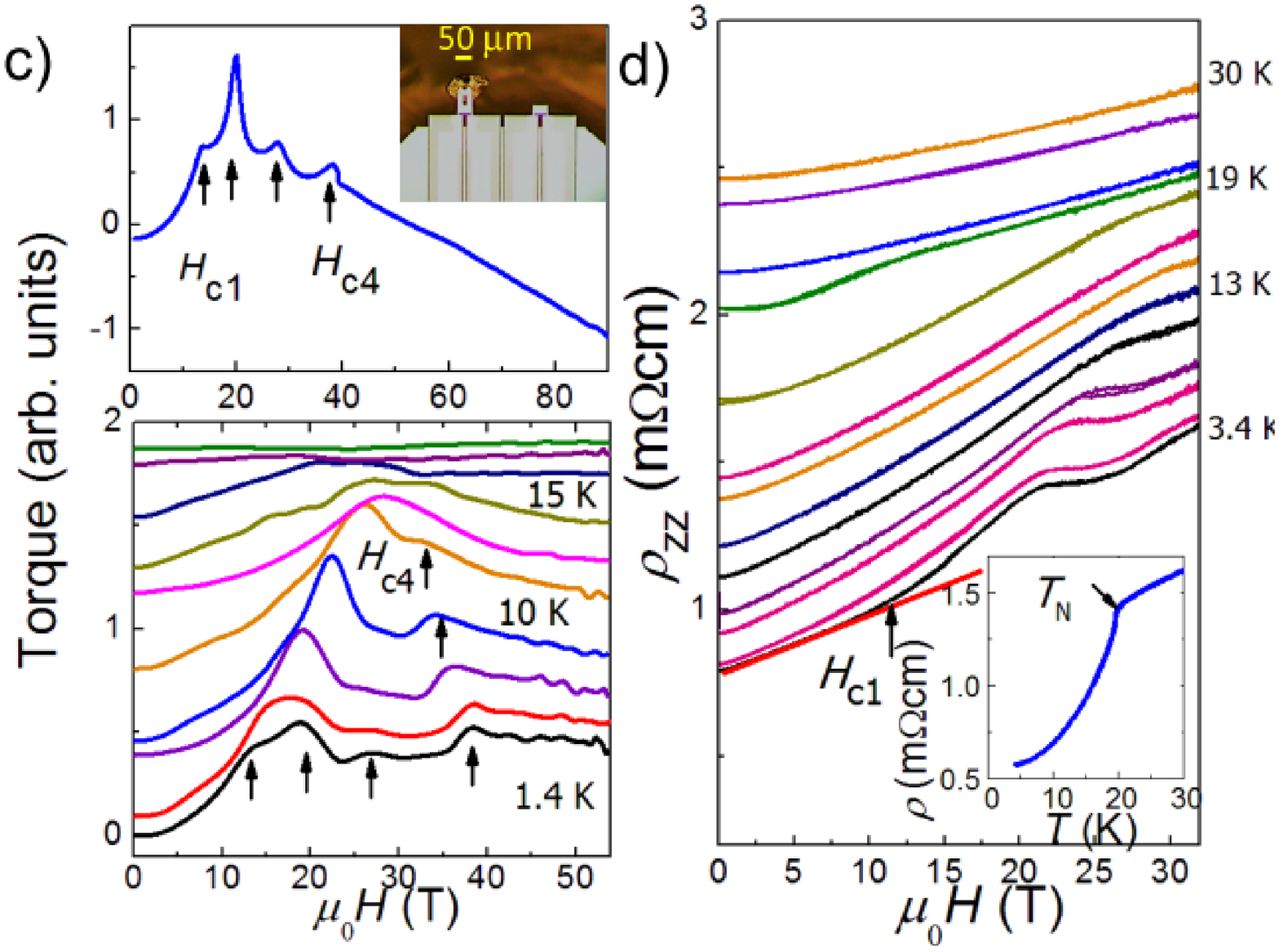}
\caption{(color online) High magnetic field measurements of
micron-size single crystals of $2H$-AgNiO$_2$.
Field dependence of (a) torque, and (b) magnetization at constant
temperature $T$ for field $\bm{H}$ nearly along the $c$-axis
($\theta \approx 3^{\circ}$).
The inset shows magnetization data for a powder sample at 5~K.
c) Torque data up to 90~T measured at constant temperatures in
pulsed magnetic fields on two different samples (top and bottom
panels). Magnetic transitions are indicated by arrows and labels.
Top inset shows a typical crystal mounted on a piezolever.
(d) Field dependence of interlayer transport at constant
temperatures below 30~K when ${\bm H}
\parallel c $ (within $\theta \approx 3^{\circ}$). The arrow
indicates the deviation from the linear dependence at $H_{c1}$.
The inset shows the low-temperature resistivity and the arrow
indicates the position of the magnetic ordering transition at
$T_{\rm N}=19.5$~K. In all panels traces at different temperatures
are uniformly shifted vertically for clarity.
}
\label{torque}
\end{figure}

For this study we use hexagonal-shaped single crystals (typical
size $\sim 70 \times 70 \times 0.1~\mu$m$^3$) grown under high
oxygen pressure \cite{growth}.
We performed a series of torque measurements (on more than 10
single crystals) using piezo-resistive, self sensing cantilevers,
at low temperatures (0.3~K) both in static magnetic fields (up to
18~T in Oxford and Bristol, 33~T at the HMFL in Nijmegen) and in
pulsed fields (up to 55~T at the LNCMP,
Toulouse and up to 90~T at NHMFL in Los Alamos).
The longitudinal magnetisation was measured by force magnetometry
using a highly sensitive magnetometer developed in Nijmegen
\cite{McCollam2011}.
%
Specific heat was measured using a purpose built micro-calorimeter using
dc and relaxation techniques. The residual resistivity ratio
is up to $\sim 250$, which indicates the high purity of the single crystals.


Magnetic torque in magnetic materials is caused by anisotropy,
measuring the misalignment of the magnetization with respect to an
uniform applied field.
The torque exerted on a sample in an applied magnetic field ${\bm
H}$ is ${\bm \tau}={\bm M}  \times \mu_0 {\bm H}$ where ${\bm M}$
is the bulk magnetization.
If ${\bm M}$ and ${\bm H}$ lie in the ($ac$) plane, then $\tau=
\mu_0 (M_a H_c- M_c H_a ) = \tfrac{1}{2} \mu_0(\chi_a -\chi_c)
H^2$ $\sin2\theta$, with $\theta = 0$ when ${\bm H} \parallel c$.
Thus, torque experiments measure the anisotropy of the
magnetization in the $ac$ plane and the torque vanishes in field
along the $c$- and $a$-axes (when $\sin2\theta=0$); the
longitudinal magnetization provides access to the parallel $M_c$
component of the sample magnetization.


Figs.~\ref{torque}(a) and (b) show the field and temperature
dependence of the torque and magnetization respectively, performed
with the magnetic field aligned close to the easy axis $c$.
At low temperatures and in low magnetic fields, the torque signal
varies as $\tau \sim H^2$ implying a constant anisotropy,
$\chi_a-\chi_c$, in the CAF phase.
By increasing the field, we observe kinks in torque at $\mu_0
H_{c1-4}$=13.5, 20, 28 and 39~T [see Figs.~\ref{torque}(a) and
(c)], which we attribute to field-induced transitions.
No further anomalies are detected at higher fields up to 90 T [see
Fig.~\ref{torque}(c)], however the torque is finite and increases
in absolute magnitude indicating that the magnetization is not yet
saturated and the region above $H_{c4}$ is most likely a phase
with spontaneous magnetic order.
Further evidence for the phase transitions seen in torque data is
provided by magnetization measurements shown in
Fig.~\ref{torque}(b). At low fields the magnetization has a weak
linear field dependence and at $H_{c1}$ the slope suddenly
changes, suggesting a linear increase in the $M_c$ component in
this phase, followed by a decrease in slope above $H_{c2}$ and a
small kink at $H_{c3}$.
Experiments of torque and specific heat in constant magnetic field
as a function of temperature presented in
Fig.~\ref{comparison-field}(a) and (b) also show clearly the
anomalies at $T_N$ and $T_{c1}$. Later, we compare in detail these
measurements with predictions for a spin Hamiltonian.

\begin{figure}[htbp]
\centering
\includegraphics[width=8.5cm,clip=true]{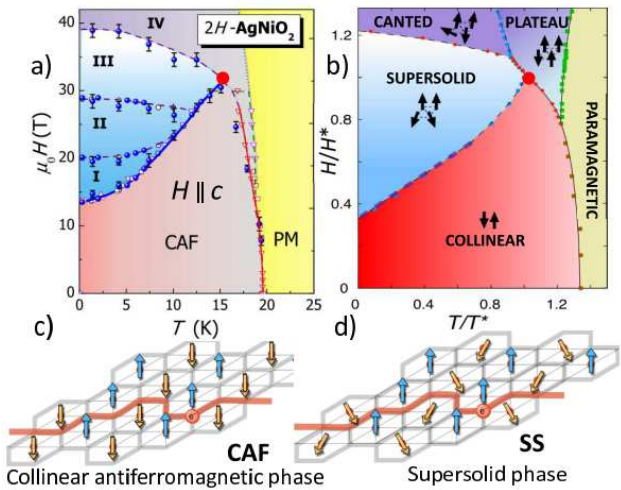}
\caption{
(colour online)
(a) The $H-T$ phase diagram of $2H$-AgNiO$_2$ from torque
magnetometry (circles), specific heat (triangles) and transport
(square).
The solid and dashed lines indicate boundaries between
different magnetic phases: collinear antiferromagnetic (CAF),
field induced phases I-IV and paramagnetic (PM).
b) The phase diagram of the classical Heisenberg model on the
triangular lattice with first- ($J_1$=$1$) and second-neighbour~($J_2$=$0.15$)
in-plane interactions, coupling between layers ($J_\perp$=$-0.15$) and easy-axis
anisotropy $D=0.25$, obtained from Monte Carlo simulation \cite{Seabra2011}.
The axes units are scaled to the point (large solid red circle)
where the CAF, SS and plateau phases meet with $T^\ast=0.3J_1$,
$B^\ast=1.6 J_1$.
In the CAF phase (c) the spin excitations are gapped, while in the SS phase (d)
electrons can scatter from gapless spin excitations.}
\label{phase_diagram}
\end{figure}

\begin{figure}[b!]
\centering
\includegraphics[width=8cm,clip=true]{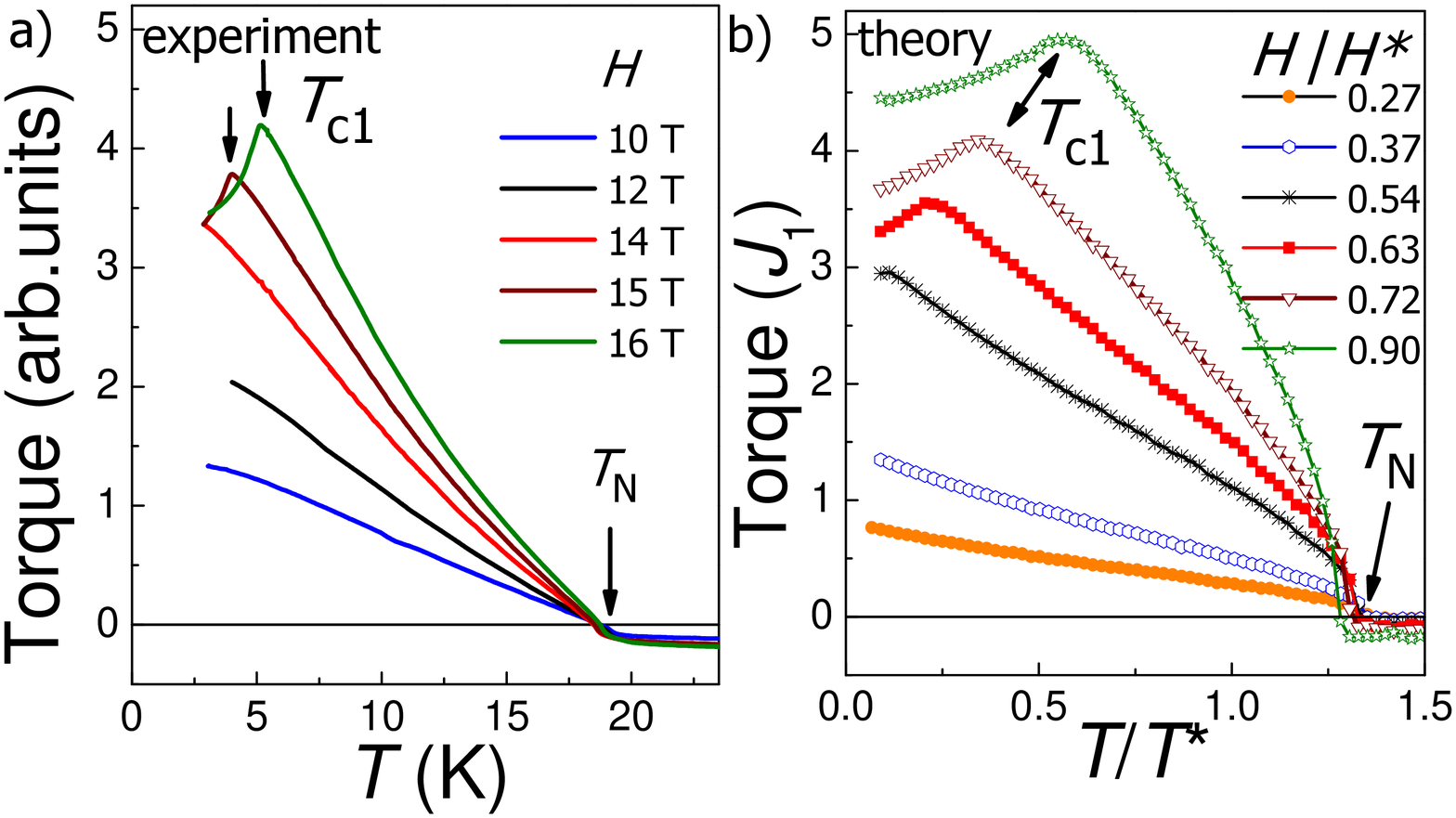}
\includegraphics[width=8cm,clip=true]{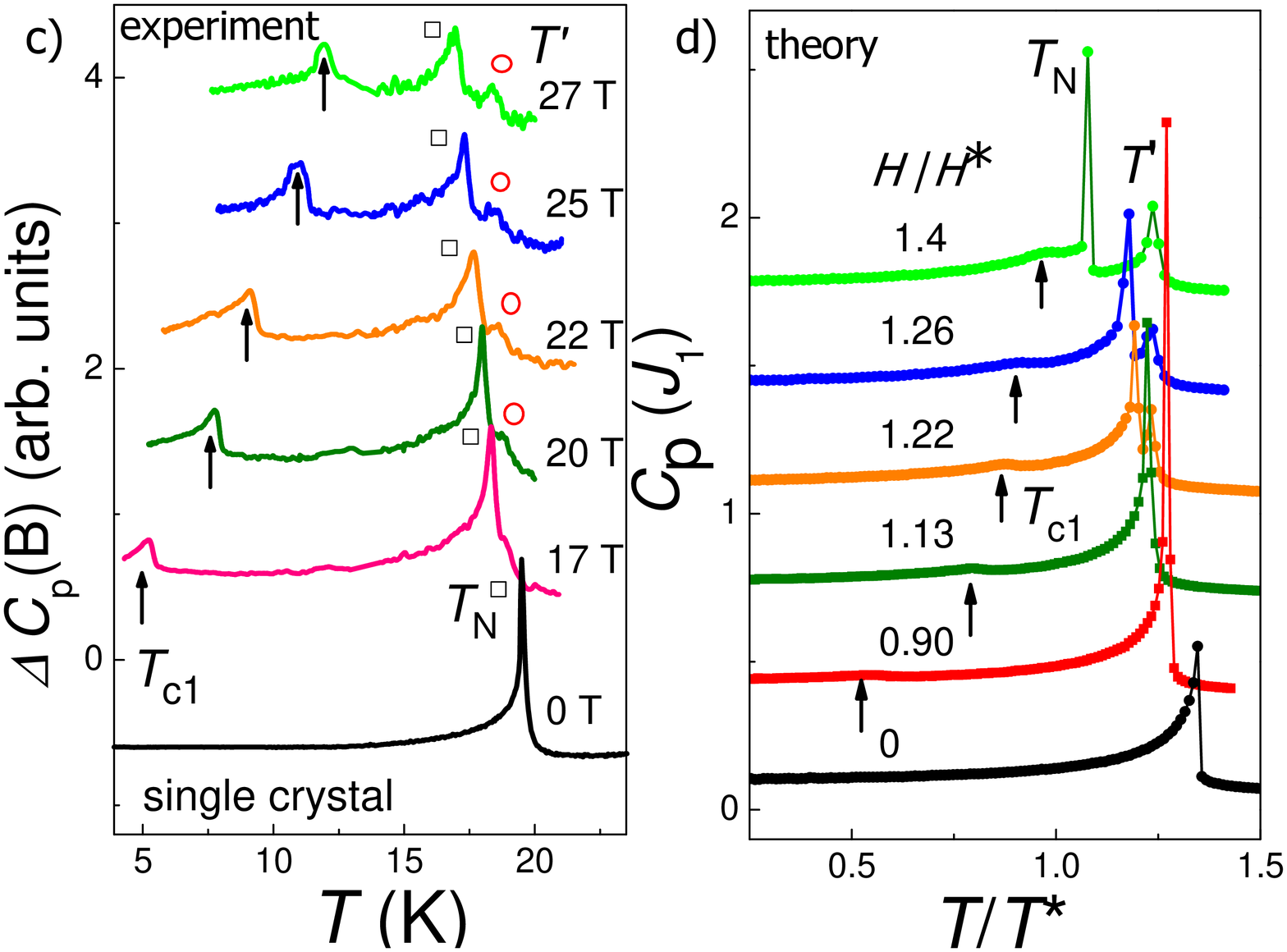}
\includegraphics[width=8cm,clip=true]{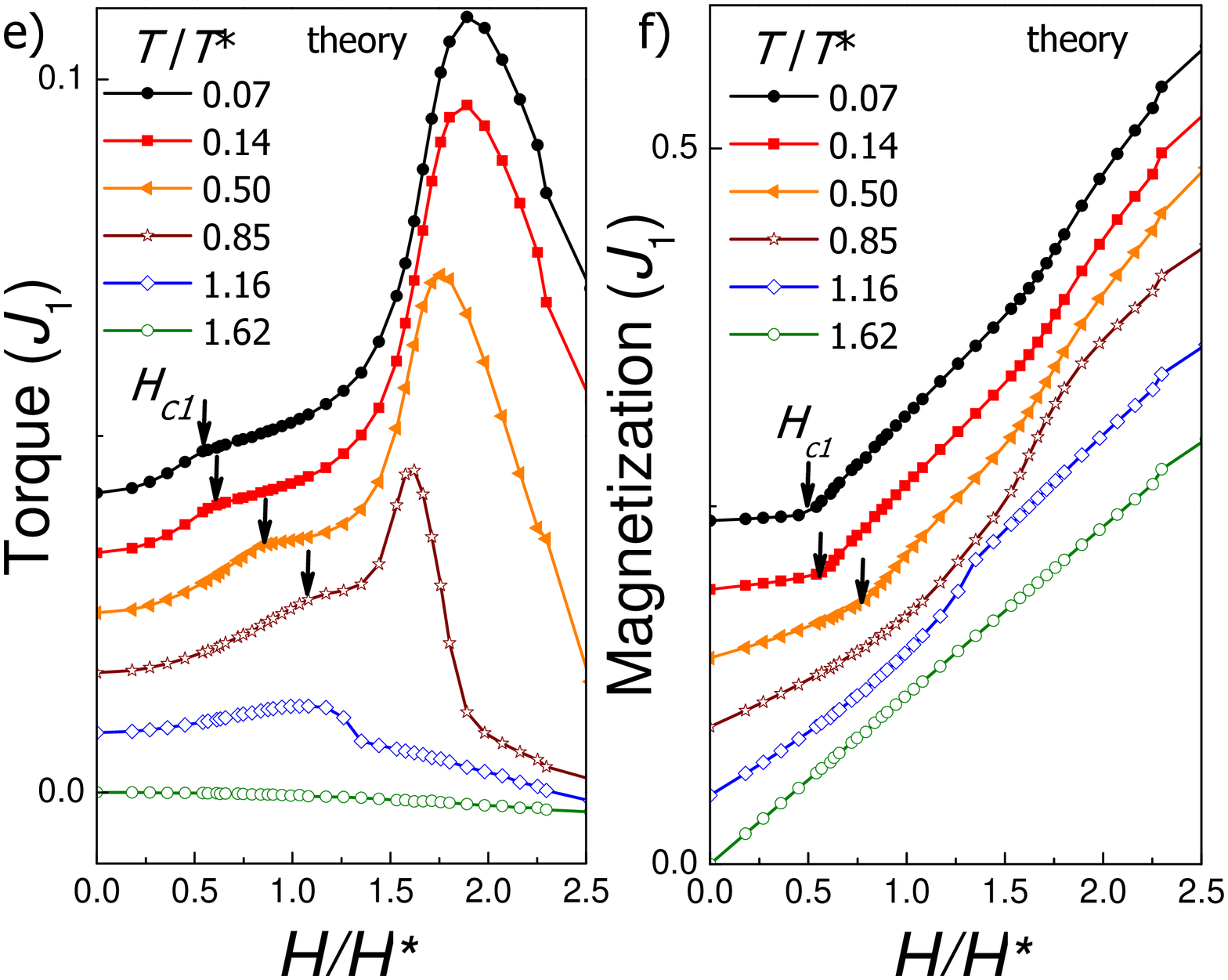}
\caption{(color online)
Temperature-dependence of torque at fixed applied field in (a)
experiments and (b) theoretical model.
Temperature-dependence of specific heat at fixed field in (c)
experiment and (d) theory.
Identified transitions are indicated by arrows at $T_{c1}$,
squares at $T_N$ and circles at $T'$.
Calculated (e) torque and (f) magnetisation as a function of
applied field at constant temperatures compared with experimental
data in Fig.\ref{torque}(a),(b) ($\theta=5^{\circ}$). In the
theoretical model saturation occurs for $H/H^\ast \approx 4.8$
with $T^\ast$ and $H^\ast$ defined in Fig~\ref{phase_diagram}(b).
Traces at different fields (c,d) and different temperatures (e,f)
are shifted vertically for clarity.
\label{comparison-field}
}
\end{figure}

Another important fact about $2H$-AgNiO$_2$ is that it is a good
metal with low residual resistivity (57~$\mu \Omega$ cm) (see
Fig.\ref{torque}(d)) and quantum oscillations have been observed
\cite{Coldea2014}.
There is a significant contributions to the density of states at
the Fermi level originating from the Ni sites on the honeycomb
lattice \cite{Wawrzynska2007}.
Fig.\ref{torque}(d) shows that that transport measurements also
exhibit anomalies at the magnetic phase transitions, showing that
the itinerant $d$ electrons are a sensitive probe of the magnetic
ground state.
There is a significant drop in resistivity below $T_{\rm N}$ (see
inset in Fig.\ref{torque}(d)), which is likely the result of
suppression of electronic scattering by low-energy spin
fluctuations when a spin gap opens below $T_{\rm N}$
\cite{Wheeler2008}.
Furthermore, magnetoresistance measurements in
Fig.~\ref{torque}(d) indicate that in the vicinity of the magnetic
transition there is a clear change in slope at $H_{c1}$ that fades
away with increasing temperature.
In zero field the Ni$^{2+}$ spins order in a collinear
antiferromagnetic pattern with spins pointing along the easy
$c$-axis, schematically shown in
Fig.~\ref{phase_diagram}(c)~\cite{Wawrzynska2007,Wheeler2008}. In
magnetic fields applied along the easy axis a transition is
expected in a field of $\Delta/(g \mu_B)$ that matches the
zero-field anisotropy spin gap, $\Delta$. Using the observed value
of the first transition field $\mu_0 H_{c1}=13.5$~T in Fig.~1(a)
gives $\Delta=1.57$~meV (using $g=2$), in good agreement with the
value of $1.7(1)$~meV estimated from inelastic neutron scattering
measurements~\cite{Wheeler2008}. For easy-axis antiferromagnets
with un-frustrated interactions the transition in field is to a
spin-flop phase, signalled by an anomaly in
torque~\cite{Nagamiya1955,Uozaki2000,Bogdanov2007}.
This canted phase is then stable with increasing magnetic field,
up to full magnetization saturation. However, for easy-axis
triangular lattice antiferromagnets with frustrated interactions,
as is believed to be the case for $2H$-AgNiO$_2$, an alternative
scenario with a richer phase diagram has been proposed
\cite{Seabra2010}, which we describe below.
%

%
The magnetic field-temperature phase diagram of $2H$-AgNiO$_2$, based on magnetic torque,
transport and specific heat data obtained over a large range of fields and
temperatures on different single crystals, is shown in Fig.~\ref{phase_diagram}(a).
Unexpectedly, in this metallic magnet, we observe a cascade of phase transitions
suggesting the formation of different magnetic structures
with increasing magnetic field,
different from typical uniaxial antiferromagnets, see e.g.
Refs.~\cite{Becerra1988,Uozaki2000,Kawamoto2008,Toft-Petersen2012},
In insulators, field-induced transitions were observed previously
in two-dimensional CuFeO$_2$ ($S=5/2$), with a related delafossite-type structure
\cite{Terada2007}, and extended models were developed  for $S=1$
\cite{Sengupta2007}.

To gain insight into the magnetism of $2H$-AgNiO$_2$ in high
magnetic fields, we consider a simple effective spin model
describing only the localised $S=1$ (Ni$^{2+}$) spins
interacting via a Heisenberg model including first- and
second-neighbour antiferromagnetic exchange ($J_1,J_2$) on the
triangular lattice, a coupling between layers ($J_{\perp}$) and
easy-axis anisotropy ($D$), using parameters obtained from fits to
powder inelastic neutron scattering data \cite{Wheeler2008}.
The resulting magnetic phase diagram in easy-axis field obtained
from classical Monte Carlo simulations was described in detail in
Refs.
~\cite{Seabra2010,Seabra2011} and in the Supplementary Material.
Here, we focus on the low-field region of the phase diagram of the spin model in
Fig.\ref{phase_diagram}(b), and we compare directly the measured and the calculated
thermodynamic quantities: torque, magnetization and specific heat.
%

The experimental phase diagram in Fig.~\ref{phase_diagram}(a)
shows that the phase boundary between the CAF phase and phase I
have an unusual field-temperature dependence, i.e. the transition
field $H_{c1}$ increases strongly with increasing temperature.
This behavior is well reproduced by the theoretical phase diagram
in Fig.~\ref{phase_diagram}(b), there the phase transition CAF-SS
is understood in terms of Bose-Einstein condensation of magnons
within the CAF state~\cite{Seabra2010}; this converts the
two-sublattice CAF order into an unusual four-sublattice state, in
which two sublattices have spins ``up'' and the other two have
spins ``down'' and canted away from the easy axis, as illustrated
in Fig.~2(d).
These canted spin components break spin-rotation symmetry about
the magnetic field, and behave like a superfluid order parameter.
Meanwhile the components of spin along the magnetic easy axis
break the discrete translational symmetry of the lattice, in a way
analogous to a solid,
therefore, the resulting state is a {\it magnetic supersolid}
\cite{Seabra2010,Seabra2011}.

Next, we compare directly the experimental results for magnetic
torque, magnetization and specific heat with those predicted by
The temperature-dependence of the torque at fixed field has a very
similar shape and qualitative form between in the experiment and
theoretical model [see Fig.~\ref{comparison-field}(a) and (b)]: in
both cases upon cooling from high temperatures in the paramagnetic
phase the torque changes sign below $T_N$, increases upon
decreasing temperature in the CAF phase, then has a peak at
$T_{c1}$, identified with the transition into the SS phase.
The temperature-dependence of the specific heat data in constant
magnetic fields also shows consistent behavior between experiment
and theory, see Fig.~\ref{comparison-field}(c) and (d). At low
fields a single anomaly is observed at the $T_{\rm N}$ transition.
In fields above $H_{c1}$ a second anomaly is observed at low
temperatures $T_{c1}$ identified with the transition CAF-SS and
this shifts to higher temperatures upon increasing
field \cite{specific_heat}. At those higher fields an additional
anomaly (labelled as $T'$) appears near $T_N$, this feature is
also present in the theoretical model, where it is associated with
the transition to another intermediate-temperature phase (labelled
``plateau'' in Fig.2(b)).

Following the same approach, we discuss the measured field
dependencies of magnetic torque and magnetisation at constant
temperature, shown in Fig.~\ref{torque}(a,b), with those predicted
by the model in Fig.~\ref{comparison-field}(e,f).
At very low temperatures in the CAF phase, the measured and
simulated magnetization are both linear in $H$, whereas the torque
shows a quadratic dependence, $H^2$, as a function of magnetic
field, and displays a kink at the first critical field, $H_{c1}$,
that becomes sharper upon lowering temperature.
Furthermore, inside the SS phase at fields right above the
transition $H_{c1}$, magnetic torque is observed to be almost
independent of the applied field (see Fig.3(e)).
All these observations are at odds with standard spin-flop transitions, see
Refs.~\cite{Nagamiya1955,Becerra1988,Uozaki2000,Bogdanov2007,Toft-Petersen2012};
in those cases, torque should show a strong divergence at a
spin-flop transition field $H_{c}$, then becomes strongly
suppressed above $H_{c}$, whereas the magnetization would have an
abrupt jump at $H_{c}$, indicative of the first-order nature of
this transition.
The transition field for a spin-flop transition is usually
independent of temperature, since it is determined by a balance of
energies between different spin configurations rather than
entropy.

While the classical spin model for the localized Ni$^{2+}$ moments
can provide a good description of many of the qualitative and
quantitative features of the lowest field-induced transition
observed at $H_{c1}$, the model cannot capture the full phase
diagram and in particular cannot account for the presence of
multiple phases spanning relatively narrow field ranges above
$H_{c1}$ (labelled as I-III).
It may be possible that itinerant electrons, neglected in the spin
model, could affect the phase diagram in this region of
intermediate fields and potentially lead to additional transitions
at $H_{c2}$ and  $H_{c3}$ \emph{inside} the SS phase of the
classical model. In the SS phase the broken translational symmetry
of the {\it solid} may lead to reconstruction of the Fermi
surface, while the gapless Goldstone modes of the {\it superfluid}
may lead to the inelastic scattering of electrons and thus an
increase of resistivity above $H_{c1}$, as observed in Fig.~1(d).
A possible reconstruction of the Fermi surface at the low-field
transition may also explain why more entropy is released in
experiment than in theory at the SS transition, indicated by the
large anomaly in specific heat at $T_{c1}$ in
Fig.~\ref{comparison-field}(c) and (d).

The observed $H-T$ phase diagram of $2H$-AgNiO$_2$
reflects the complexity of its magnetic interactions.
Since localized magnetic moments are embedded in a metal,
they are subject to RKKY interactions, which, unlike superexchange,
decays slowly with distance and may provide non-negligible further-neighbour exchange.
Furthermore, the band structure calculations show a small, but
finite magnetic moment ($m_i \approx 0.1-0.2 \mu_B$)
\cite{Wawrzynska2007} on the itinerant and inherently nonmagnetic
Ni sites on the honeycomb (see Fig.\ref{phase_diagram}c).
The Hund's rule coupling on these sites may provide an additional
incentive for the localized spins to order
and the scale of this interaction given by
$I m_i^2$/4 is a few meV
(the Stoner factor is $I \approx 0.6 - 0.8 $~eV).
By a similar mechanism, the Hund's energy of induced moments
generates ferromagnetism in SrRuO$_3$ \cite{Mazin1997}.


In conclusion, we have probed the magnetic phase diagram of the
frustrated antiferromagnetic metal, $2H$-AgNiO$_2$, in strong
magnetic fields applied along the easy axis and have observed a
cascade of magnetic phase transitions.
Thermodynamic measurements have been compared with predictions of
an effective localized spin model, which explains part of the
phase diagram and identifies a novel magnetic supersolid phase.
However, a more realistic model for 2$H$-AgNiO$_2$ needs to
consider also the itinerant $d$ electrons on the honeycomb
lattice, which may participate in the exchange interactions
and may affect the magnetic order of the localized moments.
Therefore, the itinerant electrons
may be responsible for some of the higher-field transitions observed
both in transport and thermodynamic measurements. Further studies
will explore how those phase transitions correlate with changes of
the Fermi surface topology in this frustrated magnetic metal where
$d$ electrons have mixed localized and itinerant character.

We acknowledge and thank J. Analytis, C. Jaudet, P.A. Goddard, Jos
Perenboom, M.D. Watson, for technical support during experiments.
We thank I.I. Mazin, A. Schofield, I. Vekhter for useful discussions.
This work was supported by
EPSRC Grants
EP/I004475/1,
EP/ C539974/1,
EP/G031460/1,
FCT Grant No. SFRH/BD/27862/2006,
and the EuroMagNET II
(EU Contract No. 228043).
AIC acknowledges an EPSRC Career Acceleration Fellowship (EP/I004475/1).
RMcD acknwledges support from BES "Science of 100~T".




\begin{thebibliography}{Bibliography}
\bibitem{specific_heat}
In the limit of our experimental resolution (specific heat
experiments are performed on extremely small samples) as a
function of temperature (experiments were made on
$70\times70\times10 \mu {\rm{m}}^3$ crystals) we cannot detect
clear signatures for the other two transitions clearly observed
both in the torque and transport measurements.
%
This could be due to the fact that the scans were performed mainly
at constant temperature and transition boundaries are nearly flat
and/or that these transitions are related to a reorientation of
spins which does not imply a significant change in energy and do
not result in a large anomaly in the specific heat.
\end{thebibliography}
\end{document}